# An information security monitoring and management system for 5G and 6G Networks based on SDN/NFV

Igor Buzhin, Senior Lecturer, Moscow Technical University of Communications and Informatics, Candidate of Technical Sciences (email: i.g.buzhin@mtuci.ru).

Veronica Antonova, Head of Department, Moscow Technical University of Communications and Informatics, Candidate of Technical Sciences, Associate Professor (email: xarti@mail.ru).

Yury Mironov, Dean of the Faculty, Moscow Technical University of Communications and Informatics, Candidate of Technical Sciences (email: i.b.mironov@mtuci.ru).

Vladislav Gnezdilov, Engineer, Moscow Technical University of Communications and Informatics (email: khanimood@gmail.com).

Eldar Gaifutdinov, postgraduate student, Moscow Technical University of Communications and Informatics (email: e.a.gaifutdinov@mtuci.ru).

Mikhail Gorodnichev, Dean of the Faculty, Moscow Technical University of Communications and Informatics, Candidate of Technical Sciences (email: m.g.gorodnichev@mtuci.ru).

*Abstract* - **An approach to using the concept of Software-Defined Networking and Network Functions Virtualization (SDN/NFV) for the implementation of an information security monitoring and management system in 5G and 6G networks is proposed. SDN switches based on the OpenFlow protocol are offered as network sensors. In order to reduce the time for finding a subset of the right rules in the vast array of all rules on traffic filtering systems that are logically located on sensors, a method of processing and filtering traffic in 5G and 6G transport networks is proposed. This method is based on DPDK with the LPM algorithm and is capable of processing up to 8 megapackets per second on 1 CPU core; the packet processing takes O(1), which is significantly lower than with similar algorithms. The managing subsystem consists of regional monitoring centres and a main one. The main Monitoring Centre includes a main cluster of SDN controllers along with Active/Active redundancy scheme. The regional centres represent SDN software controllers that manage locally subordinate sensors. All the managing centres are interconnected via the Transport subsystem and form a network. An algorithm for network sensor load balancing between SDN controllers has been developed in order to provide fault tolerance, load balancing and network connectivity. The algorithm results in a set of optimal sensor groups with total load not exceeding the maximum capacity of the SDN controllers.**

## I. Introduction

g collection of data about computer network operations in order to support such functions as searching for slow, faulty, or vice verse underutilized systems as well as major network resource consum-ers, meeting the parameters of SLA (Service Level Agreement) and the quality of the services provided (e.g. delivering audio, video and other content on demand, delays in connecting distributed and integrated systems). In addition to the technical items listed above, monitoring also implies supervising network performance that means controlling compliance with access, information exchange and routing policies. Collected data can reflect various aspects of network operating according to the purpose and objectives of monitoring. Both of the proposed meanings of the term "monitoring" are close to each other and rely on the same methods of data collection and analysis.

Conventional solutions for communication network monitoring consider traffic analysis based on SNMP [1]. SNMP and NetFlow [2] /IPFIX / sFlow (Sampled Flow) [3] developed much later are used to monitor communication networks as well. In the NetFlow and IPFIX protocols, flow records are made on the basis of network flows with the same attributes such as the source interface, source and destination IP, TCP/UDP and IP ToS source and destination port. While being processed these packet fields are retrieved and a hash function is computed for them to allow searching for the existing record for the flow in the shared cache. Periodically or after the timeout period, the flow records are sent to the flow collector for traffic analysis. In NetFlow, such flow records present the number of active host-to-host connections. Unlike NetFlow/IPFIX, the sFlow protocol selects packet headers and sends this information to be analyzed while not recording the flow for active connections and flushing flow caches. As a result, trans-mission delay is significantly reduced compared to NetFlow since monitoring information for selected packets is immediately available for analysis.

Monitoring of transferred data in 5G and 6G networks (Core Nework, Midhaul, Backhaul) by means of NetFlow/IPFIX [5],[9],[10] introduces a significant measurement delay because measurement data is not reported until the connection being measured is completed, which makes the length of the delay proportional to the connection time. The sFlow packet sampling method introduces significantly lower latency because sampled packets are immediately available for traffic analysis and therefore allow large flows to be detected quickly. For this reason, sFlow is appropriate for anomaly detection in Software-Defined Networking (SDN) [4] environments, although its sampling-based procedure provides less accuracy if not enough packets are selected. The development of SDN and NFV makes it possible to apply the OpenFlow (OF) protocol [6],[7] for the purposes of monitoring primarily data networks using SDN and NFV (Core Nework, Midhaul, Backhaul) [17]. The prime method of using is integration with well-known conventional monitoring tools that transfer the attributes collected by OF to NetFlow or IPFIX. An approach should be defined to take advantage of the OF protocol and the equipment that supports it in order to monitor traditional networks.

g and filtering rules in large networks is ipfirewall [23], an open source unit ported to many operating systems. Ipfw is a user-defined utility to manage ipfirewall. This utility is used to interact with the kernel module. The features of rule processing in ipfw are the rule processing from the top down, the importance of rule order, the possibility of optimizing the operation through trees and hash tables. Another solution for filtering rules is iptables/nftables that are Linux utilities providing filtering and classification of network packets/datagrams/frames. The rules are combined into chains, which in turn are included into tables. The tables contain chains. The user defines the number of tables and their headings. However, each table has only one addressing set and is only applied to packets of this set. Tables can belong to one of five sets. A chain contains rules. When running, these utilities check the rules in the chain one by one from the top down. Moreover, since rules have a certain execution "cost", it is not recommended to change their order based only on empirical observations about the byte/packet counter. The inability to specify several different actions in one rule makes iptables/nftables utilities as well as Ipfirewall low in performance and not suitable for processing a large number (about 30,000) of filtering rules. To ensure the traffic processing speed (100G+), versa-tility of writing filtering algorithms, independence from a hardware platform vendor the implementation of a filtering system for 5G and coming 6G transport network re-quires the development of an algorithm to reduce the time for finding filtering rule matching.

The application of SDN for traffic filtering and methods of optimizing rule finding are considered in [20],[21],[22], [24]. These papers propose transferring unknown packets received by the switches along preplanned routes using a "default rule". This approach can reduce the time to find the right rule for a packet, but the "default rule" may direct the flows along a nonoptimized path or block it. In [25], it is suggested to minimize the number of controlling messages by aggregating rules. The given approach will also reduce the search time for the right rule, but in this case, the controller will have ag-gregated information about the data transmission level, which may lead to QoS degradation and increased information security threats. A solution is needed to optimize OpenFlow switch rules without changing the way they work.

Thus, it is necessary to describe an approach for the implementation of an information security monitoring and management system for 5G and future 6G networks as a distributed hierarchical scheme. In order to reduce the lookup time for filtering rules in this approach, a way of processing and filtering traffic in 5G and 6G transport networks has to be proposed. To ensure fault tolerance, load balancing, network connectivity of the information security monitoring and management system, a sensor load-balancing algorithm between managing devices is to be developed.

The article is organized as follows: the introduction discusses the principles of organizing monitoring in communication networks, examines the shortcomings of monitoring methods and suggests possible methods for their elimination, discusses the principles and features of building communication networks based on SDN-NFV technologies; Section 2 analyzes the architecture of the information security monitoring and management system based on SDN-NFV technology; Section 3 proposes a method for processing and filtering traffic in 5G and 6G transport networks, comparing the developed method with well-known algorithms; in section 4, an algorithm for balancing the load coming from SDN switches between SDN controllers is proposed to ensure fault tolerance and distribution of the control loop of the monitoring system; in conclusion, the main conclusions are given.

### 1.1. Main features of the Software-Defined Networking concept

The SDN/NFV technology presents a set of techniques that enable software-based management and control of network resources usage (loading), which simplifies the task of ensuring efficient use of communication network bandwidth and this network scaling, contributes to reducing operating costs by centralizing and automating management functions. The SDN architecture [4],[5] contains three levels (Fig. 1): network infrastructure, management level and application level. SDN requires an SDN controller in the network, which provides applications with an abstract representation of network resources and ensure orchestration (coordination) of network resource management. This approach allows the controller to have access to the global state of the network and make decisions about forwarding network traffic, while the hardware is only responsible for actual transmitting information to its destinations according to the controller's instructions (packet processing rule sets).

Thus, management functions of the switch are given to a separate central device that is an SDN controller [6],[7]. Such approach allows managing and monitoring the state of a network by means of a logically centralized controller. In addition, logical representation of a network as a whole makes it possible to separate the management level from the physical portion. Interoperability between the data transmission level is carried out through a single unified open interface. SDN brings the network architecture of transport networks close to the possibility of using NFV which provides the transfer of software networking functions to cloud storage [8], and these functions are performed on servers (including virtual machines) at data processing centers (DPC).

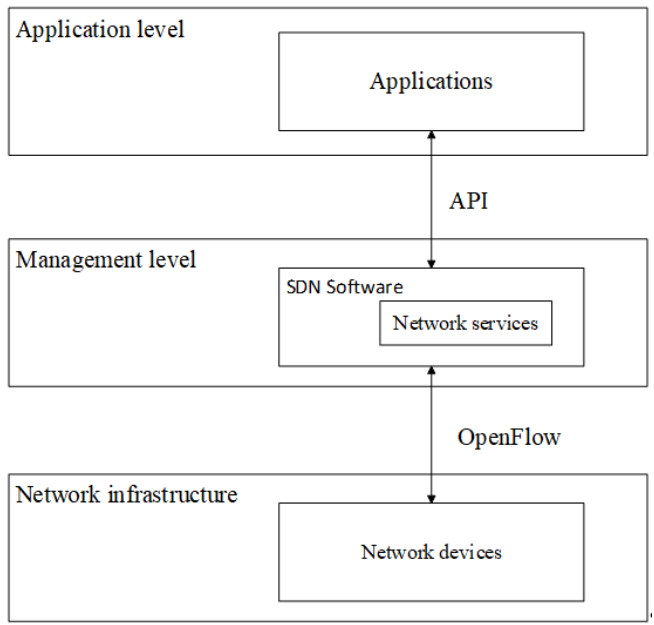

Fig. 1. Architecture SDN

In accordance with the NFV reference architecture [4],[8],[11], a set of software and hardware tools for network functions virtualization can be simplified to consist if the following basic elements (Fig. 2):

- virtual networking functions (VNF(s))
- network function virtualization infrastructure (NFVI)
- management software and orchestration NFV.

VNF(s) functions replace the ones that are performed by communication hardware in conventional communication networks in which virtualization is not applied. The network function virtualization infrastructure (NFVI) includes hardware tools and software ones that provide an abstract representation of the hardware tools the functions of which are virtualized. The NFV orchestrator (NFVO) ensures the coordinated functioning of the virtual infrastructure. The VNF manager carries out the life cycle management of the virtual network function. One manager is able to coordinate several virtual networking functions. The VIM manager performs management and control functions that provide interoperability between VNF functions and related hardware resources including the identification of hardware and software tools such as hypervisors, allocation of virtual machines, resource reallocation between virtual machines, data collection to monitor failures and capacity utilization, and so on.

Given the above features, the basic components of software-defined networks based on the OpenFlow protocol are OpenFlow Switch [13], Controller [18] or Network Operating System, Switch Management Protocol, Network Equipment Administration Protocol, Secure Channel which carries out the communication between the Controller and the Switch using the OpenFlow protocol [9],[10].

The switch is equipped with a set of addressing flow tables that form an addressing pipeline, which consists of one or more sequentially connected addressing tables [14],[15]. A packet arrived at one of the switch's input ports is first processed (service information from the packet is read), then it arrives at the addressing pipeline. The sequential pro-cessing of the packet in the addressing tables begins. A packet here means a bit string which can be divided into two parts: the header and the payload. The operations per-formed on packet flows in the addressing tables do not change the packet payload, but they can change its header.

Flow records correspond to the packets in order of priority with the first appro-priate record used in each table. When Flow Tables are applied, neither switching (L2 forwarding) nor routing (L3) occurs. We need a new term that will define this mecha-nism and that is Flow Forwarding. The meaning of the term implies that L2 and L3 searching requires much processing. The forwarding network device does not assume that the destination point for a group of identical packets can be the same. Every packet is to be assessed for making a good decision about its forwarding. Flows are processed in another way. They contain information about the state. They have the information to help the device take a decision in complex situations. Such decisions are not limited by searching based on MAC or IP destination address. They can rely on labels, VLAN ID or other identification information. This can be the information, such as DNS information, that has been agreed to other devices by the application. This approach allows making a single decision about the transfer for the whole flow and applying it very quickly.

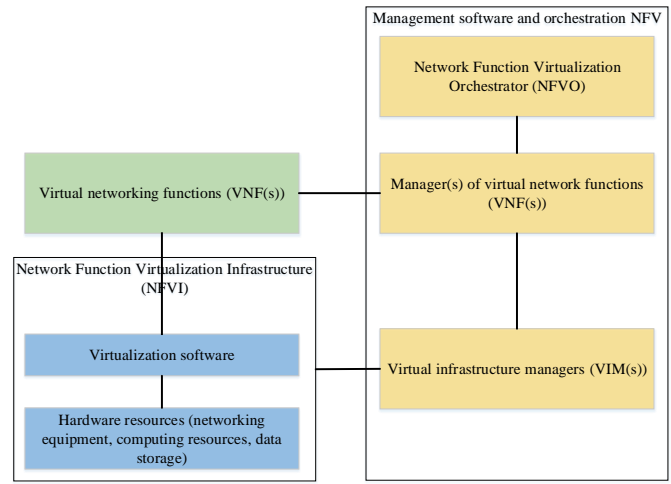

Fig. 2. Simplified Network Function Virtualization Reference Architecture

Thus, the capabilities of OpenFlow in conjunction with appropriate controller ap-plications make it possible to collect and accumulate all the data which used to be col-lected by means of IPFIX, NetFlow or sFlow protocols. This flexibility of the SDN coupled with the open software interfaces gives clear advantages in the development of network monitoring capabilities.

## II. The architecture of an information security monitoring and management system based on SDN/NFV

Рассмотрим варианты реализации системы мониторинга и Consider options for implementing an information security monitoring and man-agement system (ISMMS) in the form of a fully centralized scheme and in the form of a distributed scheme. In the fully centralized scheme all the functions of filtering, classification, data transmitting and storing as well as their management are carried out from the single center. In the distributed hierarchical scheme the ISMMS functionality and management are dispersed across all the automatic resources complexes of the distrib-uted ISMMS system for the purpose of scalability and redundancy of the latter.

The use of SDN allows the ISMMS to be implemented as a distributed hierarchical scheme. The SDN technology enables to flexibly organize the hierarchy and subordi-nation of subsystems, processing and storage servers, their levels and the degree of subordination. Moreover, the key features of this approach include the ability to dy-namically manage both the organization of the hierarchy and subordination, and the degree of redundancy and fault tolerance of its subsystems and components. In order to meet the requirements listed above the ISMMS is to be comprised of the following subsystems:

• Filtering, Classification and Storage Subsystem

• ISMMS Management Subsystem

• Transport Subsystem – the system of data transmission between sensors, regional data processing and storage centers (RDCs) and main data processing and storage center (MDC).

SDN integrated to the ISMMS allows using IDS and IPS methodology and capabilities as well as applying the analysis of the received data to centrally reprogram the network for the purpose of repelling malicious attacks and restoring functionality. This can make SDN significantly more resilient to various faults, failures and malicious at-tacks than traditional networks.

One of the key elements of the ISMMS is Filtering, Classification and Storage Subsystem. A sensor, in which the primary analysis that is network data filtering occurs, is considered to be the source of information for monitoring objectives. The sensor is an ISMMS component which is aimed at sniffing the transport network monitored by the ISMMS; hereinafter it is referred to as the monitored transport network (MTC). In order to reduce the load on the ISMMS transport network, it is reasonable to assign the filtering functions to the sensor. It will cut off unnecessary for further analysis MTC traffic from the ISMMS transport network and reduce the load on both the transport network itself and on the ISMMS storage and analysis resources.

Applying the SDN concept for the organization of the filtering subsystem is con-sidered to be most appropriate. The filtering function involves the selection of packets circulating in the MTC based on the values of packet header fields at L2-L4 levels. Building the sensor according to the SDN principles, namely in the form of the SDN switch operating on the basis of the OpenFlow protocol, will make it possible to elim-inate the problems related to a large amount of buffer storage and a large number of network interfaces. A sensor of such kind will enable dynamic traffic filtering and forwarding it to the specified ports in compliance with the rules of filtration. All of the above is a rationale for the justified application of the SDN approach for the sensor building. Thus, the basic component of the considered monitoring network is proposed to be a sensor implemented on the basis of the SDN switch supported by OpenFlow.

The storage system in the ISMMS requires the hierarchical structure and is built according to the regional principle, i.e. there are RDCs and MDC connected by the ISMMS transport network (Transport Subsystem). RDCs perform the function of intermediate buffering and temporary bandwidth shortage smoothing from sensors to RDCs; in addition, the RDC control loop can take part in the filtering subsystem management. Information in every RDC and MDC should be coherent and consistent. This implies that MDC building requires the distributed data network storage (DDS) technologies

MMS is comprised of the MDC's control loop, RDCs' control loops and the control loop of the transport network that provides the data transfer from sensors through the RDCs to the MDC, delivers com-mands for controlling network devices of the ISMMS data loop and carries out the ser-vice information delivery to the ISMMS. The ISMMS control loop is to be fault –tolerant and to ensure continuous access to RDC and MDC services. An essential aspect of the distributed nature of the ISMMS control loop is efficient scalability of the monitoring system when the scale of the data transmission network monitored by the ISMMS changes. Another important aspect of the distributed structure of the ISMMS control loop is to improve its security, namely to isolate the data transmission network between the ISMMS control loop components from the outside world so that it can be attacked only from within. The ISMMS management subsystem represents a multilevel net-working model (Fig. 3) consisting of:

• Main ISMMS Monitoring Center located at the MDC (a major cluster of SPN controllers)

• Regional ISMMS Monitoring Centers located at RDCs.

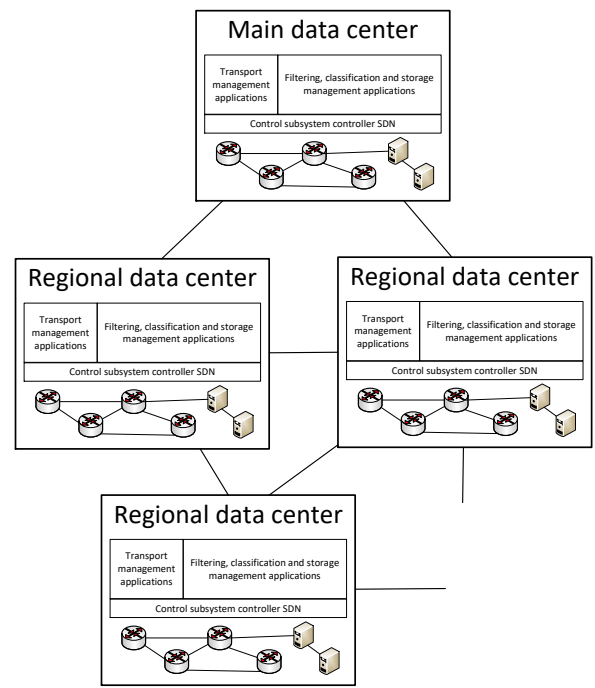

Fig. 3. Structure of the management subsystem of the monitoring system and information security management system

The Main ISMMS Monitoring Center and Regional Monitoring Centers are connected to each other by means of the Transport subsystem and compose a network. For the purpose of fault tolerance, load balancing and network connectivity it is necessary to provide both vertical and horizontal links between all the RDCs and between the RDCs and the MDC as shown in the figure. The Management subsystem organizes logically the subordination of RDCs and MDC according to the adopted ISMMS management policy flexibly building the logical subordination through software configuration of channels to provide the interconnection of RDCs and MDC. In case of emergencies such as breakdown of some RDCs, disruption of physical connections used in the configured data flow model or sudden resource shortage in an RDC or in communication lines, the Management subsystem is able to reconfigure dynamically the routs and to distribute the load over the network ensuring the compliance with the required level of service for the entire ISMMS.

Thus, the Main ISMMS Monitoring Center includes the major cluster of SDN con-trollers that is a logically centralized but physically distributed controller. This cluster operates under the scheme of Active/Active redundancy, i.e. a separate supervising SDN controller is allocated to every regional center (one or several), and in case of the controller failure the region's management is transferred to other operating controller from the major SDN controller cluster. Every major cluster's controller responsible for a region is an application for a cluster of regional center controllers.

### III. Traffic processing and filtering in 5G and 6G transport networks

Processing and filtering of external traffic at the network perimeter is an important task of a traffic filtering system. This solution is to be a software one that is in-stalled and used on standard hardware network solutions. But the number of hardware devices is limited, and if scaled, the growth in the number of devices will be non-linear. At the same time, the implementation of traffic processing and filtering at the network perimeter is a responsible process: providing there is an error in the filtering rules, partial or complete failures of the internal network can occur. In this case, the total speed of traffic processing and filtering should be 300 Gbit/s or higher.\

There are situations where complex rules increase to the order of 30000, and where networks with the so-called Wildcard mask, which is a mask that shows which part (how many bits) of the IP address can be changed, appear in the rules. The mask can be used in routing protocols such as IGRP, EIGRP, OSPF and access lists. The principle of its operation is the same as that of a normal mask except that zeros are substituted for ones and ones for zeros. Writing algorithms to handle such number of rules and the ones with such features is a complex and difficult process.

To meet the requirements of traffic processing speed (100G+), universality of writing filtering algorithms, independence from the hardware platform manufacturer, it is recommended to implement a filtering system for 5G and coming 6G transport net-works by using the DPDK technology [26], in which the interaction with the network card is carried out through specialized drivers and libraries. Incoming packets get into a ring buffer. The application regularly checks this buffer for new packets. If there are new packet descriptors in the buffer, the application accesses the DPDK packet buffers in a dedicated memory pool via pointers in the packet descriptors. If there are no packets in the ring buffer, the application polls the network devices under DPDK control and then accesses the ring again.

When DPDK is utilized for traffic processing and filtering, Data Plane is consid-ered to be separated from Control Plane (Fig. 4). The task of Data Plane `ıs to transfer packets from a queue to a queue and to modify them. Control Plane is responsible for Data Plane's dynamic configuration and control (logic port configuration, MAC address definition, IP addresses definition, VLAN configuration, BGP route configuration, ACL rule compilation in an optimal way, collection and aggregation of Data Plane's internal state metrics and tables).

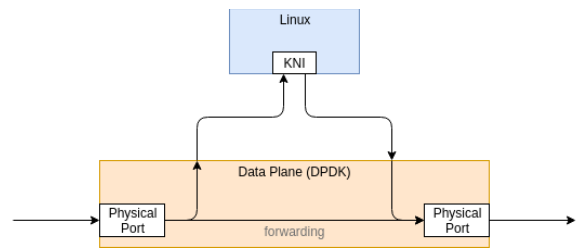

Fig. 4. Пример использования DPDK

The challenge is that ACL in 5G and 6G transport networks can have a large number of rules, and when a packet arrives at a filter it is needed to find a necessary match regardless of the size of the rule set and preserving the logic of the rule order (the problem of finding the subset of the necessary rules in a huge set of all rules).

The contents of ACL rules can be classified into the following fields: Protocol, Source Port, Destination Port, Source Net, Destination Net. Thus, several initial subsets are distinguished at the first stage. Further, if we renumber the bit masks in each classifier in order and multiply all paths to reach the final rules, we get the Cartesian product of all sets of classifiers, and its power (and computation power) will be too high. The solution is to partition the original sets into pairs or triples to reduce the power of intermediate sets and to decrease the time of searching for matches. This approach corresponds to the recursive flow classification (RFC) algorithm [19] and is shown schematically in Fig. 5 and 6.

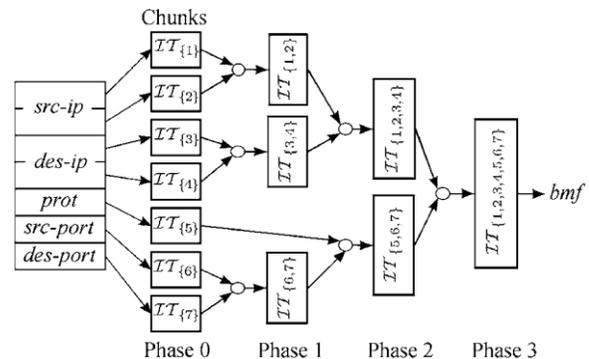

Fig. 5. Reduction tree of RFC for a typical IPv4 5-tuple classifier.

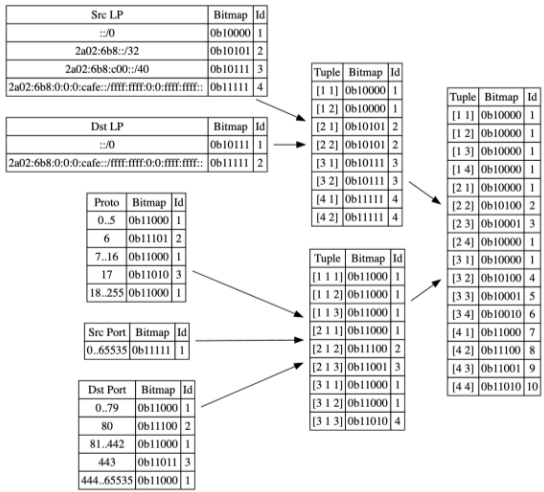

Fig. 6. Поиск соответствий в подмножествах с помощью RFC

RFC [30] transfers most of the complex operations to the preprocessing stage to build efficient data structures for high-speed classification. For a classifier of the dimension k (k is a constant), RFC requires a constant number of memory accesses to classify a packet, i.e. its temporal classification complexity is O(k). However, the complexity of its storage and preprocessing is rather high, which justifies using RFC with a large amount of memory.

In order to save prefix match operation (ACL rules match operation) time complexity there are the DPDK LPM and LPM6 libraries. These library components implement the Longest Prefix Match (LPM) table search methods for 32-bit keys (in case of LPM for IPv4 forwarding) and 128-bit keys (in case of LPM for IPv6 forwarding) [27].

A binary trie and path-compressed (a.k.a radix) trie are basic trie-based algorithms for LPM. As known these trie-based algorithms take linear time [28], so their time com-plexity is O(k), where k - prefix length. As the demand for faster LPM methods in-creased along with the interface speed, optimized variants of trie-based algorithms were proposed.

The design and working principle of LPM are the same as Trie data-structure [29]. When a key is passed for the lookup operation, the key is broken down into bytes or words and then each word (or byte) will be used one by one to traverse over the data-structure to get the required value. The time required to do a lookup operation in tries depends on the number of pieces into which a key is broken. It does not depend on the number of entries present in the tries. Asymptotically, Time Complexity of Lookup = O(m) where m is the number of bytes/words into which key is broken. If the length of the key is always constant, then we can say that lookup in trie takes O(1) or constant time [27].

One of those optimized variant of trie-based algorithm is DIR-24-8, that underlies the DPDK LPM library implementation. The main idea of DIR-24-8 algorithm is the basic scheme - DIR-24-8-BASIC - it makes use of the two tables shown in Figure 7 [23]. The first table (called TBL24) stores all possible route prefixes that are up to, and including, 24-bits long. This table has $2^{24}$ entries, addressed from 0.0.0 to 255.255.255. The second table (TBLlong) stores all route prefixes in the routing table that are longer than 24-bits. Each 24-bit prefix that has at least one route longer than 24 bits is allocated $2^8 = 256$ entries in TBLlong. Each entry in TBLlong corresponds to one of the 256 possible longer prefixes that share the single 24-bit prefix in TBL24.

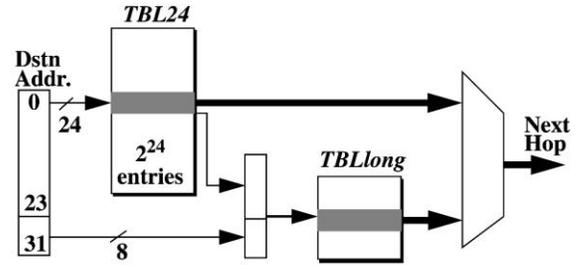

Figure 7. DIR-24-8-BASIC architecture.

The current DPDK LPM library implementation uses a variation of the DIR-24-8 algorithm that trades memory usage for improved LPM lookup speed [21]. The algo-rithm allows the lookup operation to be performed with typically a single memory read access with constant time complexity O(1). In the statistically rare case when the best match rule is having a depth bigger than 24, the lookup operation requires two memory read accesses with the same constant time complexity for each memory read access. Therefore, the performance of the LPM lookup operation is greatly influenced by whether the specific memory location is present in the processor cache or not.

The main data structure in this library implementation is almost the same as the DIR-24-8-BASIC architecture with the exception of TBLlong (tbl8 in DPDK) tables number. It is the result of conclusion that since every entry of the TBL24 (tbl24 in DPDK) can potentially point to a tbl8, ideally, it should be $2^{24}$ tbl8s, which would be the same as having a single table with $2^{32}$ entries. This is not feasible due to resource restrictions. Instead, this approach takes advantage of the fact that rules longer than 24 bits are very rare. Greatly reduction of memory consumption while maintaining a very good lookup speed (one memory access, most of the times) can be achieved by splitting the process in two different tables/levels and limiting the number of tbl8s.

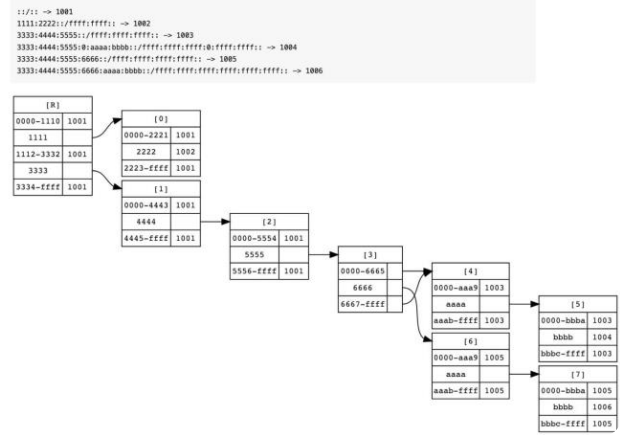

Fig. 8. An example of looking up rules in ACL using LPM6.

Figure 8 illustrates the example of finding rule matching in ACL using LPM6. As a result, the implementation of ACL on

DPDK using the LPM algorithm processes up to 8 megapackets per second on one CPU core.

A comparative analysis of the complexity of the algorithms is presented in Table 1.

Table 1. Comparison of complexity of filtering algorithms.

| Algorithm | Time complexity |
|---|---|
| Common ACL | $O(k*l)$ |
| RFC | $O(k)$ |
| LPM | $O(1)$ |

In conclusion, DPDK LPM algorithm's lookup operation can reach constant time complexity in case of prefix length is less than 24 bits or required tbl8 is in cache. Oth-erwise, there is some sort of cache miss, and in that case lookup operation requires two memory read accesses. This case happens statistically rare, so average lookup operation time complexity is $O(1)$. Thereby, with expected time complexity $O(1)$ LPM algorithm is more computational efficient than RFC algorithm with time complexity $O(k)$ in case of prefix lookup operation (e.g. ACL rule matching).

IV. The algorithm for load balancing in the information security monitoring and management system based on SDN/NFV

An essential aspect of the distributed nature of the ISMMS control loop is to ensure its fault tolerance. The distribution of the ISMMS control loop is convenient in terms of providing resource abundance that can be used if one or several components fail for some reasons. The functionality of faulty resources can be reallocated among the other ones, if necessary. Thus, the problem of balancing the load received from SDN switches (ISMMS sensors) between SDN controllers arises. The load on the SDN switches is always changing during the operation of the ISMMS, which may cause the SDN controller to lack the capacity to process messages from the switches it manages. In this case it is necessary to carry out load balancing of the SDN controllers by transferring switch management from one controller to another one. In this situation, the load represents the Packet-In messages received by the controller per second. The messages are generated by switches at the time every new flow arrives. The controller processes the message data and sets rules for every switch in the segment. Thus, it is necessary to allocate the SDN switches to the controllers so that the minimum possible number of Packet-In messages is generated.

The OpenFlow protocol versions 1.3.3 and later introduced a tool for separating controller roles in relation to the switch. It describes three SDN controller operation modes: Master, Slave and Equal. In the Slave mode the controller has a read-only access to the switch and does not get Packet-In messages. The Master and Equal modes allow the controller to manage switches and receive Packet-In messages from them. Changes in controller operation modes are made by means of "Role Change" messages. To implement the switch management redistribution between various controllers the algorithm from [5] can be applied.

To distribute the load (Packet-In messages received by the controller per second) from the SDN/NFV based ISMMS sensors among the SDN controllers, all SDN switches should be divided into groups depending on the load they produce so that the total load from one group of switches does not exceed the performance capacity of the SDN controller. This requires ongoing measurements of loading at every controller. The principal task is the load balancing problem itself, which is the allocation of SDN switches (network ISMMS sensors) to their controllers. The task needs to be divided into two subtasks, each one running according to its own algorithm. The first algorithm should apply the information about the current loads and flows in the system to divide the switches into groups, so that the resulting groups can then be assigned to the controllers existing in the system. The second algorithm should distribute the groups resulting from the first algorithm to the existing controllers in the way the minimum number of migrations is required to transform the existing system. The developed algorithm for distributing Packet-In messages arriving at controllers has no analogues for comparison.

To improve the efficiency and optimize the performance of networks, including 5G, graph theory and hypergraph theory are often used [22]. To implement the first algorithm we build an undirected graph $G = (S, E)$; its vertices correspond to the sensors existing in the system and its edges correlate with the communication channels used between sensors. Here $|S|$ is the number of sensors (SDN switches) in the network (numbered 0 through $|S| - 1$), i.e. $S = \{S_i, i = 0, \dots, |S| - 1\}$ is a set of SDN switches (note that a switch is a network node which can host an SDN controller in addition to the SDN switch); $E = \{E_{i,j} : E_{i,j} = (S_i, S_j)\}$ is a set of connections between network nodes (SDN switches); $|E|$ is the number of connections between network nodes. $C = \{c_i, i = 0, \dots, |C| - 1\}$ is also known and equals a set of SDN controllers; $|C|$ is the number of SDN controllers of the network. Let $P_i$ be the maximum capacity (the number of Packet-In messages per second) of the controller $c_i$. Let $I_i$ be equal to the average number of new flows arriving at the switch $S_i$ per second, and let $I_{ij}$ be the average number of flows per second between the switches $S_i$ and $S_j$. It is necessary to divide the graph $G = (S, E)$ into the connectivity groups so that the sum of the numbers in the nodes of each connectivity group does not exceed the controller's capacity $P_i$. Therefore, the problem of dividing switches into groups is the task of splitting the graph into connectivity components. The objective for the second stage is to divide the graph into such connectivity components where each connectivity component satisfies the condition that the sum of the numbers in the nodes of this component does not exceed the controller capacity.

$I_i + I_{ij}$ is calculated on the nodes of every connectivity component. If $I_i + I_{ij} \leq P_i$, the algorithm completes its operation.

If $I_i + I_{ij} > P_i$, it is necessary to:

1. Build a new graph with vertices $S$ without connections: $G^1 = (S, E^{нов})$.

2. Choose the connection $E_{i,j}$ with maximum $I_{ij}$ in the graph $G$ and delete it. Add to the graph $G^1$. Loading is recalculated on the nodes of the graph $G$ as follows: the loads corresponding to the endpoints of the edges that have been removed $G$ are subtracted from the existing load values. The division into groups is obtained.

3. If $I_i + I_{ij} \leq P_i$ at each of the connectivity component of the new graph $G$, the integration was successful. If $I_i + I_{ij} > P_i$, the integration is considered to fail. The next (by load) edge of the graph $G$ is deleted.

As a result of the algorithm a set of optimal groups of switches with the total load not exceeding the maximum capacity of the SDN controllers is acquired.

Furthermore, it is necessary to assign the obtained optimal groups of SDN switches to the controllers C so that bringing the system to the state corresponding to the distribution into the groups would require as few actions as possible. For this purpose the groups of SDN switches are to be sorted in a descending order of the number of loads in each switch group. The resulting sorted set of switch groups is compared to the current allocation of switches to SDN controllers (switches managed by the same controller are counted in one group). Based on the comparison of the two sets, we perform reconnection (migration and changing the controller's modes) of switches to controllers only in places of mismatch of the two sets. After performing these actions a new allocation of switches to controllers is obtained.

Thus, there are three stages in the algorithm. The network configuration can be presented as a graph, in which the nodes are SDN switches (ISMMS sensors) with the rated load about the network states and information security events, and the controllers with the rated capacity of the data flow processing; the edges are data transmission channels. The first step is to remove the edges that have minimal load on the interaction channel of two switches until each sum of the values of the connectivity component nodes of the graph is less than the capacity of the controller. At the same time, if the removal of an edge forms a new connectivity component, the load equal to half the load of the removed channel is added to the rated value of the nodes between which the removal is performed. Otherwise, if no component is formed, half the load of the removed channel is added to the buffer of the switches between which the edge is removed. The value from the buffer is added when a new connectivity component has appeared. Thus, when a channel is deleted, the load remained is allowed for and will travel through other communication channels. The second step is to fill the graph with the edges that have the minimum load of the number of deleted edges until such a connectivity component is formed that the sum of the load values of the terminal switches of every connectivity component does not exceed the controller capacity and a new addition of an edge will result in controller overloading. When an edge is added, half the load of the switch interaction added in the first stage is deleted from the values of the switches. The third stage is the connection of the formed connectivity components to the redundant controllers in the descending order of the sums of node values included to connectivity components, i.e. the smaller ones will be assigned to redundant controllers.

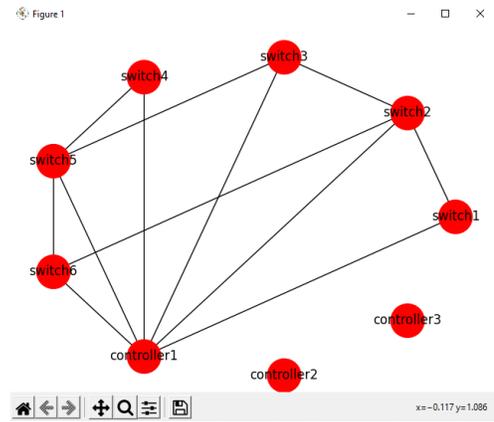

Fig. 9. An example of the functioning of the balancing algorithm: the example of the 1st stage of the balancing algorithm running.

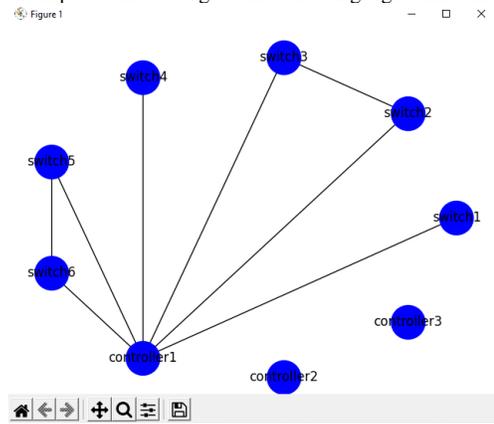

Fig. 10. An example of the functioning of the balancing algorithm: the example of the 2nd stage of the balancing algorithm running.

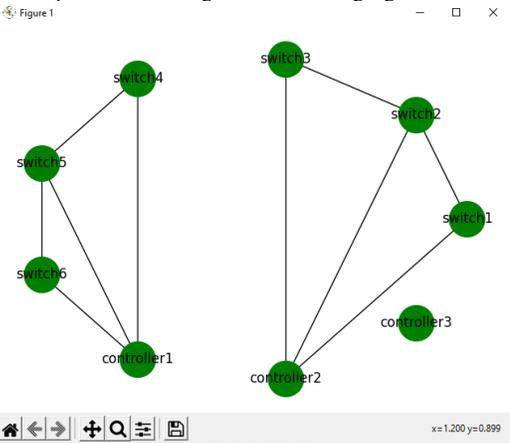

Fig. 11. An example of the functioning of the balancing algorithm: the example of the 3rd stage of the balancing algorithm running.

Fig. 9, 10, 11 shows an example of the loadbalancing algorithm in the ISMMS imple-mented in Python. Fig. 9 illustrates the allocation of six switches and three SDN controllers. At the first stage the edges between switches 4 and 5, 1 and 2, 3 and 5, 2 and 6 were deleted. Fig. 10 demonstrates the example of the second stage of the balancing algorithm. At this stage the edges between switches 1 and 2, and 4 and 5 were added. The example of the third stage of the algorithm is given in Fig. 11 Reconnection of one of the connectivity components to controller 2 is carried out at this stage.

## V. Conclusions

The paper provides the principles and architecture of the information security monitoring and management system (ISMMS) for 5G and 6G networks based on SDN/NFV. The application of SDN/NFV allows the ISMMS to be implemented in the form of the distributed hierarchical scheme. The key elements of the ISMMS are filtering, classification and storage subsystem, ISMMS management subsystem, transport subsystem that is a system of data transmitting between sensors, RDCs and MDC. The basic component of the considered monitoring network is proposed to be a sensor implemented on the basis of the SDN switch supported by OpenFlow. In order to improve the efficiency of traffic processing and filtering on 5G and 6G transport networks, a method of the implementaction of ACL on DPDK using the LPM algorithm is presented; the method allows processing up to 8 megapackets per second on one CPU core while processing the packet takes O(1), which is significantly lower than the RFC algorithm. To provide fault tolerance and distribution of the ISMMS control loop, an algorithm for balancing the load from SDN switches (ISMMS sensors) between SDN controllers is proposed. As a result of the algorithm a set of optimal groups of sensors, the total load of which does not exceed the maximum performance of SDN controllers, is obtained